\documentclass{svmult}

\usepackage{makeidx}         
\usepackage{graphicx}        
\usepackage{multicol}        
\usepackage[bottom]{footmisc}

\usepackage{amssymb}   
\usepackage{amsmath}   
\usepackage{bbm}
\usepackage{psfrag}    

\def\sfracolaf#1#2{{\textstyle\frac{#1}{#2}}}
\def\rdolaf#1{\buildrel{_{_{\hskip 0.01in}\rightarrow}}\over{#1}}
\def\ldolaf#1{\buildrel{_{_{\hskip 0.01in}\leftarrow}}\over{#1}}

\newcommand{\unityolaf}{{\mathbbm{1}}}
\newcommand{\velolaf}{\mathrm{v}}
\newcommand{\trolaf}{\mathrm{tr}}
\newcommand{\Rolaf}{{\mathbb{R}}}
\newcommand{\Colaf}{{\mathbb{C}}}
\newcommand{\Zolaf}{{\mathbb{Z}}}
\newcommand{\Nolaf}{{\mathbb{N}}}
\newcommand{\undolaf}{\qquad\textrm{and}\qquad}

\begin{document}

\title*{Noncommutative Solitons \\[8pt]
\small{150 minutes of lectures given 01-04 November 2005 at the 
International Workshop on Noncommutative Geometry and Physics
in Sendai, Japan}}
\titlerunning{Noncommutative Solitons}
\author{Olaf Lechtenfeld}
\institute{Institut f\"ur Theoretische Physik, Universit\"at Hannover,\\
	   Appelstra\ss{}e 2, D--30167 Hannover, Germany }

\maketitle


\section{Introduction}

Noncommutative geometry is a possible framework for extending our
current description of nature towards a unification of gravity with
quantum physics. In particular, motivated by findings in string theory,
field theories defined on Moyal-deformed spacetimes (or brane world-volumes)
have attracted considerable interest. For reviews, 
see~\cite{konechnyo,douglaso,szaboo1}.

In modern gauge theory, a central role is played by nonperturbative objects,
such as instantons, vortices and monopoles. These solitonic classical field
configurations usually arise in a BPS sector (or even integrable sector) of
the theory. This fact allows for their explicit construction and admits
rather detailed investigations of their dynamics.

It is then natural to ask how much of these beautiful results survives the
Moyal deformation and carries over to the noncommutative realm. Studying
classical solutions of noncommutative field theories is also important to
establish the solitonic nature of D-branes in string theory (see the reviews
\cite{harveyo,hamanakao,szaboo2}). It turns out that already scalar field 
theories, when Moyal deformed, have a much richer spectrum of soliton solutions
than their commutative counterparts.

The simplest case in point are noncommutative scalar field theories in
one time and two space dimensions. Therefore, in these lectures I will
concentrate on models with one or two real or with one complex scalar field.
In the latter case my prime example is the abelian unitary sigma model and
its Grassmannian subsectors. By adding a WZW-like term to the action it is
extended to the noncommutative Ward model~\cite{wardo1,wardo2,ioazak1,lepo1}, 
which is integrable in $1{+}2$ dimensions and features exact multi-soliton 
configurations. As another virtue this model can be reduced to various 
lower-dimensional integrable systems such as the sine-Gordon theory
\cite{lmppt}.

After covering the basics in the beginning of these lectures, I will 
present firstly static and secondly moving abelian sigma-model solitons,
with space-space and with time-space noncommutativity. Multi-soliton
configurations and their scattering behavior will make a brief appearance. 
The nonabelian generalization is sketched in a U(2) example.
Next, I shall compute the full moduli-space metric for the abelian Ward model 
and discuss its adiabatic two-soliton dynamics.
A linear stability analysis for a prominent class of static U(1)~solitons 
follows, with an identification of their moduli. 
In the last part of the lectures, I dimensionally reduce to 
$1{+}1$ dimensions and find noncommutative instantons as well as solitons. 
The latter require also an algebraic reduction, 
from U(2) to U(1)${\times}$U(1), which produces an integrable noncommutative
sine-Gordon model. Its classical kink and tree-level meson dynamics will close 
the lectures. The material of these lectures is taken from the papers
\cite{lepo1,lepo2,dolepe,chule,klalepe}.

I'd like to add that the topics presented here are by no means exhaustive.
I have deliberately left out important issues such as noncommutative vortices,
monopoles and instantons, the role of the Seiberg-Witten map, quantum aspects
such as renormalization, or non-Moyal spaces like fuzzy spheres and quantum 
groups. The noncommutative extension of integrable systems technology 
(ADHM, twistor methods, dressing, Riemann-Hilbert problem etc.) 
is also missing. Finally, I did not touch the embedding in the framework of 
string theory. Any of these themes requires lectures on its own.

\section{Beating Derrick's theorem}

\subsection{Solitons in $d=1{+}2$ scalar field theory}

I consider a real scalar field $\phi$ living at time~$t$ on a plane
with complex coordinates~$z,\bar z$. The standard action
\begin{equation}
S_0 \ =\ \int\!\D{t}\,\D^2{z}\; \bigl[ \sfracolaf12 \dot\phi^2
- \partial_z\phi\,\partial_{\bar z}\phi - V(\phi) \bigr]
\end{equation}
depends on a polynomial potential~$V$ of which I specify
\begin{equation}
V(\phi)\ge 0 \quad,\qquad
V(\phi_0)=0 \undolaf
V'(\phi) \ =\ v\prod_i (\phi-\phi_i) \quad.
\end{equation}
In this situation, Derrick's theorem states that the only non-singular
static solutions to the equation of motion are the ground states $\phi=\phi_0$,
allowing for degeneracy. The argument is strikingly simple~\cite{derricko}:
Assume you have found a static solution~$\widehat\phi(z,\bar z)$.
Then by scaling I define a family
\begin{equation}
\widehat\phi_\lambda(z,\bar z) \ :=\ 
\widehat\phi(\sfracolaf{z}{\lambda},\sfracolaf{\bar z}{\lambda})
\end{equation}
of static configurations, which must extremize the energy at~$\lambda{=}1$.
However, over a time interval~$T$, I find that
\begin{equation}
E(\lambda) \ :=\ -S_0[\widehat\phi_\lambda]/T \ =\ 
\lambda^0\,E_\textrm{grad} + \lambda^2\,E_\textrm{pot} \quad,
\end{equation}
which is extremal at~$\lambda{=}1$ only for $E_\textrm{pot}=0$,
implying $E_\textrm{grad}=0$ and $\widehat\phi_1=\phi_0$ as well.

\subsection{Noncommutative deformation}

Let me deform space (but not time) noncommutatively, by replacing the
ordinary product of functions with a so-called star product, which is
noncommutative but associative. This step introduces
a dimensionful parameter~$\theta$ into the model, which I use to
define dimensionless coordinates~$a,\bar a$ via
\begin{equation}
z \ =\ \sqrt{2\theta}\,a \undolaf \bar z\ =\ \sqrt{2\theta}\,\bar a \quad.
\end{equation}
For static configurations, the energy functional then becomes
\begin{equation}
E_\theta \ =\ \int\!\D^2{a}\; \bigl[
|\partial_a \phi|^2_\star + 2\theta\,V_\star(\phi) \bigr]
\quad\buildrel{\theta\to\infty}\over\longrightarrow\quad
2\theta\int\!\D^2{a}\;V_\star(\phi) \quad,
\end{equation}
where the subscript `$\star$' signifies star-product multiplication.
In the large-$\theta$ limit, the stationarity equation obviously becomes
\begin{equation}
0 \ =\ V'_\star(\widehat\phi) \ =\ v (\widehat\phi{-}\phi_0)\star
(\widehat\phi{-}\phi_1)\star\cdots\star(\widehat\phi{-}\phi_n) \quad.
\end{equation}
Due to the noncommutativity (you may alternatively think of~$\widehat\phi$
as a matrix) this equation has many more solutions than just
$\widehat\phi=\phi_i=\textrm{const}$, namely
\begin{equation} \label{phsololaf}
\widehat\phi \ =\ U\!\star \bigl(\sum_i \phi_i P_i \bigr)\star U^\dagger
\qquad\textrm{with}\qquad P_i\star P_j = \delta_{ij} P_j \undolaf
\sum_i P_i = \unityolaf 
\end{equation}
featuring a resolution of the identity into a complete set of (star-)projectors
$\{P_i\}$ and an arbitrary star-product unitary~$U$.
The energy of these solutions comes out as
\begin{equation}
E_\theta[\widehat\phi]\quad\buildrel{\theta\to\infty}\over\longrightarrow\quad
2\theta\sum_i V(\phi_i) \int\!\D^2{a}\;P_i \ =\
2\pi\theta\sum_{i\neq0} V(\phi_i)\ \trolaf P_i \quad,
\end{equation}
where I defined the `trace' via $\trolaf P=\pi\int\!\D^2{a}\,P$.
Clearly, the moduli space of the large-$\theta$ solutions~(\ref{phsololaf}) 
is the infinite-dimensional coset 
$\frac{\textrm{U}(\infty)}{\prod_i\textrm{U(rank}P_i)}$. 
For finite values of~$\theta$, the effect of the gradient term in the action
lifts this infinite degeneracy and destabilizes most solutions.

\subsection{Moyal star product}

Specifying to the Moyal star product, I shall from now on use
\begin{equation}
\begin{aligned}
(f\star g)(z,\bar z) 
&\ =\ f(z,\bar z)\,\exp\,\bigl\{ \theta\,
({\ldolaf{\partial}}_z {\rdolaf{\partial}}_{\bar z} -
{\ldolaf{\partial}}_{\bar z} {\rdolaf{\partial}}_z ) \bigr\}\,g(z,\bar z) 
\\[6pt]
&\ =\ f\,g\ +\ 
\theta\,(\partial_z f\,\partial_{\bar z} g-\partial_{\bar z} f\,\partial_z g)
\ +\ \ldots \\[6pt]
&\ =\ f\,g\ +\ \textrm{total derivatives}
\end{aligned}
\end{equation}
with a constant noncommutativity parameter~$\theta\in\Rolaf_+$.
The most important properties of this product are
\begin{equation}
(f\,\star\,g)\,\star\,h \ =\ f\,\star\,(g\,\star\,h) \quad,\qquad
\int\!\!\D^2{z}\ f\star g\ =\ \int\!\!\D^2{z}\,f\,g \quad,\qquad
[z\,,\,\bar z]_\star \ =\ 2\,\theta \quad.
\end{equation}

\subsection{Fock-space realization}

A very practical way to realize the Moyal-deformed algebra of functions
on~$\Rolaf^2$ is by operators acting on a Hilbert space~$\cal H$.
This realization is provided by the Moyal-Weyl map between functions~$f$
and operators~$F$, i.e.
\begin{equation}
\bigl( f(z,\bar z),\star\bigr) \quad\leftrightarrow\quad
\bigl( F(a,a^\dagger),\cdot \bigr) \quad.
\end{equation}
For the coordinate functions I take
\begin{equation}
z \quad\leftrightarrow\quad \sqrt{2\theta}\,a 
\qquad\textrm{such that}\qquad [\,a\,,\,a^\dagger]\ =\ \unityolaf \quad.
\end{equation}
The concrete translation prescriptions read
\begin{equation}
F \ =\ \textrm{Weyl-order} 
\Bigl[ f\bigl(\sqrt{2\theta}\,a,\sqrt{2\theta}\,\bar a\bigr) \Bigr]
\undolaf f \ =\ F_\star 
\bigl(\sfracolaf{z}{\sqrt{2\theta}},\sfracolaf{\bar z}{\sqrt{2\theta}}\bigr)
\quad,
\end{equation}
and derivatives and integrals become algebraic:
\begin{equation}
\sqrt{2\theta}\,\partial_z f  \ \leftrightarrow\ -[a^\dagger,F] \quad,\qquad
\sqrt{2\theta}\,\partial_{\bar z} f\ \leftrightarrow\  [a,F] \quad,\qquad
\smallint\!\D^2z\;f\ =\ 2\pi\theta\;\trolaf F
\end{equation}
where the trace runs over the oscillator Fock space~$\cal H$ with basis
\begin{equation} \label{Nbaseolaf}
|n\rangle \ =\ \sfracolaf{1}{\sqrt{n!}}\,(a^\dagger)^{n}\,|0\rangle 
\qquad\textrm{for}\qquad n\in\Nolaf_0 \undolaf a\,|0\rangle =0 \quad.
\end{equation}

\section{$d=0{+}2$ sigma model}

\subsection{$\textrm{U}_\star(1)$ sigma model in $d=0{+}2_\theta$}

To be specific, let me turn to the simplest noncommutative sigma model
in the $2d$ plane, i.e. the abelian sigma model,
\begin{equation}
\phi \in \textrm{U}_\star(1) \qquad\Longleftrightarrow\qquad 
\phi\star\phi^\dagger \ =\ \unityolaf \quad.
\end{equation}
Naively, it looks like a commutative U($\infty$) sigma model.
Restricting to static fields, the action or, rather, the energy functional is
\begin{equation} \label{Eolaf}
E \ =\ 2\int\!\D^2{z}\;\partial_z\phi^\dagger\,\partial_{\bar z}\phi
\ =\ 2\pi\;\trolaf \bigl|[a,\Phi]\bigr|^2 \quad,
\end{equation}
which yields the equation of motion (I drop the hats on $\Phi$)
\begin{equation}
0 \ =\ \Phi^\dagger\,\bigl[a\,,[a^\dagger,\Phi]\bigr] - 
\bigl[a^\dagger,[a\,,\Phi^\dagger]\bigr]\,\Phi
\ =:\ \Phi^\dagger \, \Delta\Phi - \Delta\Phi^\dagger \, \Phi \quad,
\end{equation}
thereby defining the laplacian.
This model possesses an ISO(2) isometry: the Euclidean group of rigid motions
\begin{equation}
\bigl(a\,,a^\dagger\bigr) \quad\longmapsto\quad
\bigl(\E^{\I\vartheta}(a{+}\alpha)\,,
\E^{-\I\vartheta}(a^\dagger{+}\bar\alpha)\bigr)
\qquad\textrm{for}\quad \alpha\in\Colaf 
\quad\textrm{and}\quad \vartheta\in\Rolaf/2\pi\Zolaf
\end{equation}
induces the global field transformations
\begin{equation}
\Phi\quad\longmapsto\quad 
\E^{\I\vartheta\,\text{ad}(a^\dagger a)}\,
\E^{\alpha\,\text{ad}(a^\dagger)-\bar\alpha\,\text{ad}(a)}\,\Phi \ =:\ 
R(\vartheta)\,D(\alpha)\,\Phi\,D(\alpha)^\dagger\,R(\vartheta)^\dagger \quad.
\end{equation}
The unitary transformation acts on the vacuum state as
\begin{equation}
R(\vartheta)\,D(\alpha)\,|0\rangle 
\ =\ \E^{\I\vartheta\,a^\dagger a}\,
\E^{\alpha a^\dagger -\bar\alpha a}\,|0\rangle
\ =:\ |\E^{\I\vartheta} \alpha\rangle
\end{equation}
and produces coherent states. 
Furthermore, the model enjoys a global phase invariance under
\begin{equation}
\Phi\quad\longmapsto\quad\E^{\I\gamma_0}\,\Phi \quad.
\end{equation}

\subsection{Grassmannian subsectors}

There exist unitary fields which are hermitian at the same time.
The intersection of both properties yields idempotent fields,
\begin{equation}
\Phi^\dagger\ =\ \Phi \qquad\Longleftrightarrow\qquad 
\Phi^2\ =\ \unityolaf \quad,
\end{equation}
and defines hermitian projectors
\begin{equation}
\sfracolaf12(\unityolaf{-}\Phi) \ =\ P \ =\ P^2
\qquad\Longleftrightarrow\qquad \Phi \ =\ \unityolaf-2P \quad.
\end{equation}
The set of all such projectors decomposes into Grassmannian submanifolds,
\begin{equation}
\Phi\ \in\ \textrm{Gr}(r,{\cal H}) \ =\ 
\frac{\textrm{U}({\cal H})}{\textrm{U(im}P) \times \textrm{U(ker}P)}
\qquad\textrm{with}\quad r=\textrm{rank}P=0,1,2,\dots \quad.
\end{equation}
A restriction of the configuration space to some~Gr($r,\cal H$) defines
a Grassmannian sigma model embedded in the $\textrm{U}_\star(1)$ model.
Quite generally, any projector of rank~$r$ can be represented as
\begin{equation}
P \ =\ |T\rangle\,{\langle T|T\rangle}^{-1} \langle T| 
\qquad\textrm{with}\qquad 
|T\rangle \ =\ \bigl( |T_1\rangle,|T_2\rangle,\dots,|T_r\rangle \bigr) \quad,
\end{equation}
where the column vector $\langle T|$ is the hermitian conjugate of~$|T\rangle$,
and $\langle T|T\rangle$ stands for the $r{\times}r$ matrix of scalar products
$\langle T_i|T_j\rangle$. In the rank-one case, this simplifies to
\begin{equation}
|T\rangle\ \in\ \Colaf P({\cal H})\ =\ \Colaf P^\infty\quad,\qquad
\textrm{i.e.}\quad |T\rangle\ \simeq\ |T\rangle\,\Gamma
\quad\textrm{for}\quad\Gamma\in\Colaf^* \quad.
\end{equation}
If finite, the rank~$r$, which labels the Grassmannian subsectors, 
is also the value taken by the topological charge
\begin{equation}
Q \ =\ \sfracolaf{1}{8\pi}\smallint\D^2{z}\;
\phi\star\partial_{[z}\phi\star\partial_{\bar z]}\phi \ =\ 
\trolaf\bigl( P\,a\,(\unityolaf{-}P)\,a^\dagger P\ -\ 
P\,a^\dagger(\unityolaf{-}P)\,a\,P \bigr) \quad,
\end{equation}
which may be compared to the energy
\begin{equation}
\sfracolaf{1}{8\pi} E \ =\ \trolaf\bigl|[a,P]\bigr|^2 \ =\ 
\trolaf\bigl( P\,a\,(\unityolaf{-}P)\,a^\dagger P\ +\ 
P\,a^\dagger(\unityolaf{-}P)\,a\,P \bigr) \quad.
\end{equation}

\subsection{BPS configurations}

In a given Grassmannian the energy is bounded from below by a BPS argument:
\begin{equation} \label{BPSEolaf}
\sfracolaf{1}{8\pi}E \ =\
Q\ +\ \trolaf (F^\dagger F+F F^\dagger) \ \ge\ Q 
\qquad\textrm{with}\qquad F\ =\ (\unityolaf{-}P)aP \quad.
\end{equation}
For finite-rank projectors\footnote{
For finite-corank projectors, $Q=-\trolaf(\unityolaf{-}P)$ and $E\ge-Q$.} 
I have $Q=\trolaf P$ and hence $E_\textrm{BPS}=8\pi\trolaf P$.
The energy is minimized when the projector obeys the BPS equation
\begin{equation}
(\unityolaf{-}P)\,a\,P\ =\ 0 \qquad\Longleftrightarrow\qquad
a:\ \text{im}P\ \hookrightarrow\ \text{im}P \quad,
\end{equation}
which is equivalent to
\begin{equation} \label{BPSTolaf}
a\,|T\rangle\ =\ |T\rangle\,\Gamma 
\qquad\textrm{for some $r{\times}r$ matrix}\quad\Gamma \quad,
\end{equation}
meaning that $|T\rangle$ spans an $a$-stable subspace.
By a basis change inside im$P$ one can generically diagonalize~\footnote{
In general I must allow for confluent eigenvalues, which produce Jordan cells.
For each cell, the multiple state~$|\alpha\rangle$ gets replaced with 
the collection $\bigl\{|\alpha\rangle,a^\dagger|\alpha\rangle,\dots\bigr\}$.}
\begin{equation}
\Gamma\ \to\ \textrm{diag}(\alpha_1,\alpha_2,\dots,\alpha_r) \quad,
\end{equation}
whence BPS solutions are just coherent states
\begin{equation} \label{cohTolaf}
|T\rangle\ =\ 
\bigl( |\alpha_1\rangle,|\alpha_2\rangle,\dots,|\alpha_r\rangle \bigr)
\qquad\textrm{with}\qquad 
|\alpha_i\rangle\ =\ \E^{\alpha_i a^\dagger -\bar\alpha_i a}\,|0\rangle \quad.
\end{equation}
The corresponding projector reads
\begin{equation} \label{cohprojolaf}
P \ =\ \sum_{i,j=1}^r |\alpha_i\rangle\,
\bigl( \langle\alpha_.|\alpha_.\rangle \bigr)^{-1}_{ij} \langle\alpha_j|
\ =\ U \Bigl( \sum_{k=0}^{r-1} |k\rangle\langle k| \Bigr) U^\dagger \quad,
\end{equation}
where $U$ is a unitary which in general does not commute with~$a$.
To develop the intuition, I display the Moyal-Weyl image of the basic operators
\begin{eqnarray}
|\alpha\rangle\langle\beta| \quad&\leftrightarrow&\quad
2\E^{\I\kappa}\,\E^{-\frac12|\alpha-\beta|^2}\, 
\E^{-(z-\sqrt{2\theta}\alpha)(\bar z-\sqrt{2\theta}\beta)/\theta} 
\undolaf \\[6pt]
|k\rangle\langle k| \quad&\leftrightarrow&\quad  
2\,L_k(\sfracolaf{2z\bar z}{\theta})\,\E^{-z\bar z/\theta} \quad,
\end{eqnarray}
where $L_k$ denotes the $k$th Laguerre polynomial.
Obviously, $P$ is related to a superposition of gaussians in the Moyal plane.
Note that the gaussians are singular for $\theta\to0$.

\section{$d=1{+}2$ sigma model}

\subsection{$d=1{+}2$ Yang-Mills-Higgs and Ward model}

At this stage I'd like to bring back the time dimension, 
but return to the commutative situation ($\theta{=}0$) for a while. 
The sigma model of the previous section extends to $1{+}2$ dimensions
in more than one way, but only a particular generalization yields an 
integrable theory, the so-called Ward model~\cite{wardo1,wardo2,ioazak1}.
Interestingly, its equation of motion follows from specializing the 
Yang-Mills-Higgs equations: The latter are implied by the Bogomolnyi equations
\begin{equation}
\sfracolaf12\varepsilon^{abc} (\partial_{[b}A_{c]}+A_{[b}A_{c]}) \ =\
\partial^a H + [A^a,H] \qquad\textrm{with}\quad a,b,c\in\{t,x,y\} \quad,
\end{equation}
where the Yang-Mills potential~$A_a$ and the Higgs field~$H$ take values in
the Lie algebra of~U($n$) for definiteness.
A light-cone gauge and ansatz of the form
\begin{equation}
A_t \ =\ A_y\ =\ \sfracolaf12 \phi^\dagger(\partial_t+\partial_y)\phi \undolaf
A_x \ =\ -H \ =\ \sfracolaf12 \phi^\dagger \partial_x \phi
\end{equation}
yields a Yang-type Ward equation for the prepotential $\phi\in\textrm{U}(n)$,
\begin{equation} \label{wardeqolaf}
(\eta^{ab}+k_c\,\varepsilon^{cab})\partial_a(\phi^\dagger\partial_b\phi)\ =\ 0
\quad\Longleftrightarrow\quad
\partial_x(\phi^\dagger\partial_x\phi)-\partial_v(\phi^\dagger\partial_u\phi)
\ =\ 0 \quad,
\end{equation}
introducing the metric $(\eta_{ab})=\textrm{diag}(-1,+1,+1)$, a fixed vector
$(k_c)=(0,1,0)$ and the light-cone coordinates
\begin{equation}
u\ =\ \sfracolaf12(t+y) \undolaf v\ =\ \sfracolaf12(t-y) \quad.
\end{equation}

\subsection{Commutative Ward solitons}

\begin{figure}
\centering
\includegraphics[height=5cm]{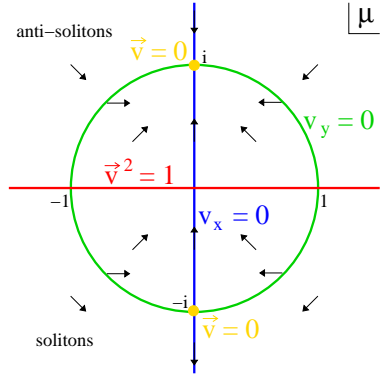}
\caption{Soliton velocities in the $\mu$ plane}
\label{fig:olaf1}       
\end{figure}

Due to the appearance of the fixed vector~$k$, the `Poincar\'e group' ISO(1,2)
is broken to the translations times the $y$-boosts. This is the price to pay 
for integrability. The existence of a Lax formulation, a linear system,
B\"acklund transformations etc.~suggest the existence of multi-solitons
in this theory, which indeed can be constructed by classical means.
Rather than directly integrating the Ward equation~(\ref{wardeqolaf}),
multi-solitons require solving only first-order equations and so in a way
are second-stage BPS solutions of the Yang-Mills-Higgs system. 
The U($n$)-valued one-soliton configurations reads
\begin{equation}
\phi \ =\ (\unityolaf{-}P)+\sfracolaf{\mu}{\bar\mu}P 
\qquad\textrm{for}\quad  \mu\in\Colaf\setminus\Rolaf \quad,
\end{equation}
with a hermitian projector
\begin{equation}
P \ =\ T\,\sfracolaf{1}{T^\dagger T}\,T^\dagger
\end{equation}
subject to
\begin{equation} \label{BPS1olaf}
(\unityolaf{-}P)\,(\bar\mu\partial_x - \partial_u)\,P \ =\ 0 
\ =\ (\unityolaf{-}P)\,(\bar\mu\partial_v - \partial_x)\,P \quad.
\end{equation}
It turns out that each finite-rank~$P$ yields a soliton with constant
velocity $(\velolaf_x,\velolaf_y)$ and energy $E$ given by
\begin{equation} \label{velEolaf}
(\velolaf_x,\velolaf_y) \ =\ -\Bigl(
\frac{\mu+\bar\mu}{\mu\bar\mu+1}\ ,\ \frac{\mu\bar\mu-1}{\mu\bar\mu+1}\Bigr)
\undolaf E\ =\ 
\frac{\sqrt{1{-}\velolaf_x^2{-}\velolaf_y^2}}{1-\velolaf_y^2}\ 8\pi\;\trolaf P
\end{equation}
making obvious the Lorentz symmetry breaking (see also figure~\ref{fig:olaf1}).

\subsection{Co-moving coordinates}

\begin{figure}
\centering
\includegraphics[height=6cm]{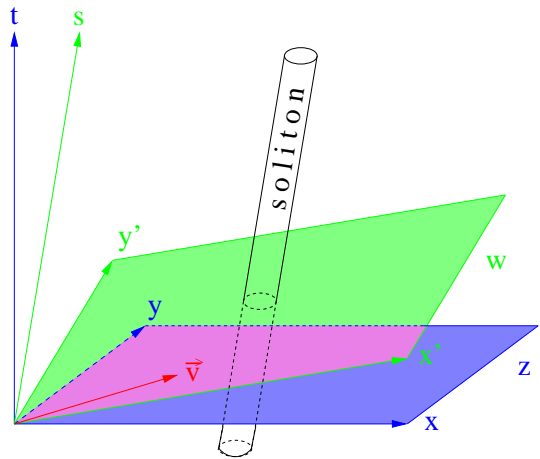}
\caption{Spacetime picture of co-moving coordinates}
\label{fig:olaf2}
\end{figure}

Since one-soliton configurations are lumps moving with constant velocity,
I can pass to their rest frame via a linear coordinate transformation
$(u,v,x)\mapsto(w,\bar w,s)$ given by
\begin{equation} \label{restolaf}
w\ =\ \nu\ \bigl[\bar{\mu}\,u+\sfracolaf{1}{\bar\mu}\,v+x\bigr]\quad,\qquad
\bar w\ =\ \bar\nu\ \bigl[\mu\,u+\sfracolaf{1}{\mu}\,v+x\bigr]\quad,\qquad
s\ =\ \dots
\end{equation}
with $\nu\in\Colaf$ to be chosen later and $s$ not needed. The transformation 
degenerates for $\mu\in\Rolaf\ \leftrightarrow\ \rdolaf\velolaf^2=1$, 
as is seen in the map for the partials,
\begin{equation}
\begin{aligned}
\partial_w &\ =\ 
\sfracolaf{1}{\nu} \sfracolaf{\mu\bar\mu}{(\mu-\bar\mu)^2}\,
\bigl[\sfracolaf{1}{\mu}\,\partial_u + \mu\,\partial_v -2\,\partial_x \bigr]
\quad,\qquad
\partial_u \ =\ 
\nu\bar\mu\,\partial_w + \bar\nu\mu\,\partial_{\bar w}
- \sfracolaf{2\I \mu\bar\mu}{\mu-\bar\mu} \,\partial_s \ \ ,
\\[6pt] 
\partial_{\bar w} &\ =\ 
\sfracolaf{1}{\bar\nu} \sfracolaf{\mu\bar\mu}{(\mu-\bar\mu)^2}\,
\bigl[\sfracolaf{1}{\bar\mu}\,\partial_u+\bar\mu\,\partial_v-2\,\partial_x\bigr]
\quad,\qquad
\partial_v \ =\ 
\sfracolaf{\nu}{\bar\mu}\,\partial_w+\sfracolaf{\bar\nu}{\mu}\,\partial_{\bar w}
- \sfracolaf{2\I}{\mu-\bar\mu}\,\partial_s \quad,
\\[6pt] 
\partial_s &\ =\ 
\sfracolaf{-\I}{\mu-\bar\mu}\,
\bigl[\partial_u + \mu\bar\mu\,\partial_v -(\mu{+}\bar\mu)\,\partial_x\bigr]
\!\quad,\qquad
\partial_x \ =\ 
\nu\,\partial_w + \bar\nu\,\partial_{\bar w}
- \sfracolaf{\I (\mu{+}\bar\mu)}{\mu -\bar\mu}\,\partial_s \quad.
\end{aligned}
\end{equation}
In the co-moving coordinates, the BPS conditions~(\ref{BPS1olaf}) reduce to
\begin{equation} \label{BPS2olaf}
(\unityolaf{-}P)\,\partial_{\bar w}\,P \ =\ 0 \ =\ 
(\unityolaf{-}P)\,\partial_s\,P \quad.
\end{equation}
The static case is recovered at
\begin{equation}
\velolaf_x=\velolaf_y=0 \quad\Longleftrightarrow\quad \mu=-\I 
\quad\Longrightarrow\quad w=\nu\,(x+\I y) \ ,\quad s=t \quad.
\end{equation}
The spacetime picture is visualized in figure~\ref{fig:olaf2}.

\subsection{Time-space versus space-space deformation}

Now I set out to Moyal-deform the Ward model.
In contrast to the static sigma model, two distinct possibilities appear,
namely space-space or time-space noncommutativity:\footnote{
I do not discuss light-like deformations here.}
\begin{equation} \label{nctypesolaf}
\begin{aligned}
{} [x\,,\,y]_\star &\ =\ \I\theta
\qquad & \Longrightarrow \qquad
& [w,\bar w]_\star\ \propto\ 
2\theta\,\nu\bar\nu\,\sqrt{\vphantom{\big|}\smash{1{-}\rdolaf{\velolaf}^2}}
\quad, \\[6pt]
{} [t\,,\,\rdolaf{n}\cdot\rdolaf{r}\,]_\star &\ =\ \I\theta
\qquad & \Longrightarrow \qquad
& [w,\bar w]_\star\ \propto\ 
2\theta\,\nu\bar\nu\,|\rdolaf{n}\times\rdolaf{\velolaf}|
\quad,
\end{aligned}
\end{equation}
where $\rdolaf{r}=(x,y)$ and $\rdolaf{n}=(n_x,n_y)=\textrm{const}$ 
in the $xy$~plane. It is apparent that the time-space deformation 
becomes singular when $\rdolaf\velolaf\parallel\rdolaf{n}$, 
including the static case $\rdolaf\velolaf=0$~!
Hence, soliton motion purely in the deformed direction yields 
{\it commutative\/} rest-frame coordinates. In all other cases, 
$(w,\bar w)$ decribes a standard Moyal plane, and each rest-frame-static
BPS projector,
\begin{equation}
\partial_s\,P\ =\ 0\undolaf(\unityolaf{-}P)\,\partial_{\bar w}\,P\ =\ 0\quad,
\end{equation}
gives a soliton solution. 
In the rest frame, the original type of deformation is no longer relevant.

\subsection{$\textrm{U}_\star(1)$ Ward solitons}

The Moyal-Weyl map associates to the co-moving coordinate~$w$ 
an annihilation operator~$c$ via
\begin{equation}
w\ \leftrightarrow\ \sqrt{2\theta}\,c \quad.
\end{equation}
Now I adjust the free parameter~$\nu$ such that
\begin{equation}
[w,\bar{w}]_\star\ =\ 2\theta \qquad\Longleftrightarrow\qquad 
[\,c\,,\,c^\dagger]\ =\ \unityolaf \quad.
\end{equation}
The BPS condition~(\ref{BPS2olaf}) becomes
\begin{equation}
(\unityolaf{-}P)\,c\,P\ =\ 0 \qquad\textrm{for projectors}\qquad 
P(w,\bar w) \ =\ |T\rangle\,{\langle T|T\rangle}^{-1} \langle T| \quad,
\end{equation}
and it is solved in the abelian case by
\begin{equation}
|T\rangle \ =\ \bigl( |T_1\rangle,|T_2\rangle,\dots,|T_r\rangle \bigr) 
\qquad\textrm{with}\qquad
|T_i\rangle\ =\ \E^{\alpha_i c^\dagger-\bar\alpha_i c}\,|\velolaf\rangle\quad,
\end{equation}
where $|\velolaf\rangle$ is the `co-moving vacuum' defined by
\begin{equation}
c\,|\velolaf\rangle\ =\ 0 \quad.
\end{equation}
One finds that the soliton velocity and energy are $\theta$~independent,
hence the commutative relations~(\ref{velEolaf}) still apply. Like in the
static case, the $\textrm{U}_\star(1)$ solitons have no commutative limit.
A change of velocity, $\rdolaf\velolaf\to\rdolaf\velolaf'$, is effected by an
ISU(1,1) squeezing transformation
\begin{equation}
c \ =\ S(t)\,c\,'\,S(t)^\dagger \undolaf 
|\velolaf\rangle \ =\ S(t)\,|\velolaf'\rangle \quad,
\end{equation}
and so all co-moving vacua~$|\velolaf\rangle$ are obtained from~$|0\rangle$
in this fashion. For the simplest case, a moving rank-one soliton, one gets
\begin{equation} \label{onesololaf}
\begin{aligned}
\Phi &\ =\ \E^{\alpha c^\dagger -\bar\alpha c}\,\bigl( \unityolaf -
(1{-}\sfracolaf{\mu}{\bar\mu})|\velolaf\rangle\langle\velolaf| \bigr)\, 
\E^{\bar\alpha c -\alpha c^\dagger} \\[6pt]
\qquad\leftrightarrow\qquad
\phi &\ =\ 1-(1{-}\sfracolaf{\mu}{\bar\mu})\,2\,
\E^{-|w-\sqrt{2\theta}\alpha|^2/\theta} \quad.
\end{aligned}
\end{equation}
Remembering that $w=w(z,\bar z,t)$ one encounters
a squeezed gaussian roaming the Moyal plane.

\subsection{Ward multi-solitons}

Integrability allows me to proceed beyond the one-soliton sector.
The dressing method, for example, allows for the construction of
multi-solitons (with relative motion). More concretely, 
a $\textrm{U}_\star(1)$ $m$-soliton configuration is built from (rows of)
states $\bigl(|T_i^{(k)}\rangle\bigr)_{i=1,\dots,r_k}^{k=1,\dots,m}$
parametrized by 
\begin{equation}
(\mu_1,\dots,\mu_m)\quad\Longleftrightarrow\quad 
(\rdolaf\velolaf_{\!1},\dots,\rdolaf\velolaf_{\!m})
\end{equation}
and $r_k{\times}r_k$ matrices $\Gamma^{(k)}$ in eigenvalue equations
\begin{equation}
c_k\,|T^{(k)}\rangle\ =\ |T^{(k)}\rangle\, \Gamma^{(k)} 
\qquad\textrm{with}\qquad 
c_k\ =\ S(\rdolaf\velolaf_{\!k},t)\ a\ S(\rdolaf\velolaf_{\!k},t)^\dagger\quad.
\end{equation}
In a basis diagonalizing~$\Gamma^{(k)}$ the solution reads
\begin{equation}
|T_i^{(k)}\rangle \ =\ |\alpha_i^k,\mu_k,t\rangle\ :=\
\E^{\alpha_i^k c_k^\dagger -\bar\alpha_i^k c_k} |\velolaf_k\rangle
\qquad\textrm{with}\qquad 
|\velolaf_k\rangle\,=\,S(\!\rdolaf\velolaf_{\!k},t)\,|0\rangle 
\end{equation}
such that $c_k|\velolaf_k\rangle=0$.
The two-soliton with $r_1=r_2=1$ provides the simplest example:
\begin{equation}
\Phi \ =\ \unityolaf\ -\ \sfracolaf{1}{1-\mu|\sigma|^2} \,\bigl[\,
\sfracolaf{\bar\mu_{11}}{\bar\mu_1}|1\rangle\langle1|\ +\ 
\sfracolaf{\bar\mu_{22}}{\bar\mu_2}|2\rangle\langle2|\ -\
\sigma\mu\sfracolaf{\bar\mu_{12}}{\bar\mu_1}|1\rangle\langle2|\ -\
\bar\sigma\mu\sfracolaf{\bar\mu_{21}}{\bar\mu_2}|2\rangle\langle1| \,\bigr]
\end{equation}
with the abbreviations
\begin{equation}
|k\rangle\equiv|\alpha^k,\mu_k,t\rangle \quad,\quad\
\sigma\equiv\langle1|2\rangle \quad,\quad\ 
\mu_{ij}\equiv\mu_i{-}\bar\mu_j \quad, \quad\ 
\mu\equiv\sfracolaf{\mu_{11}\,\mu_{22}}{\mu_{12}\,\mu_{21}} \quad.
\end{equation}
Because of the no-force property familiar to integrable models,
the energy is additive:
\begin{equation}
E[\Phi] \ =\ E(\mu_1)\ +\ E(\mu_2) \quad,
\end{equation}
with $E(\mu)$ given in~(\ref{velEolaf}) and $\trolaf P_i=r_i=1$.
The two lumps distort each other's shape but escape the overlap region
as if each one had been alone.

\subsection{Ward soliton scattering}

It follows that abelian Ward multi-solitons are squeezed gaussian lumps 
moving with different but constant velocities~$\rdolaf\velolaf_{\!k}$ 
in the Moyal plane.
The large-time asymptotics of these configurations shows no scattering
for pairwise distinct velocities. However, in coinciding-velocity limits
there appear new types of multi-solitons with novel time dependence.
This kind of behavior extends to the nonabelian case. Moreover,
$\textrm{U}_\star(n{>}1)$ multi-solitons (with zero asymptotic relative
velocity) as well as soliton-antisoliton configurations~\cite{lepo2,wolfo}
can be made to scatter at rational angles~$\frac\pi q$ 
in this manner. In addition, breather-like ring-shaped bound states
are found as well. Unfortunately, for $\textrm{U}_\star(1)$ only the
latter kind of configurations appear in the coinciding-velocity limits,
hence true scattering solutions are absent for abelian solitons.

\subsection{$\textrm{U}_\star(n)$ Ward solitons}

For completeness, let me briefly illustrate how the generalization to
the nonabelian case works. I restrict myself to a comparison of
$\textrm{U}_\star(1)$ with $\textrm{U}_\star(2)$ static one-solitons.
Since the nonabelian BPS projectors have infinite rank, it is convenient
to switch from states~$|T\rangle$ to operators $\widehat T$:
\begin{equation}
\textrm{U}_\star(1): \qquad
|T\rangle \ =:\ \widehat T\,|{\cal H}\rangle  \qquad\textrm{with}\qquad
|{\cal H}\rangle\ \equiv\ \bigl( |0\rangle\ |1\rangle\ |2\rangle\ \dots\bigr)
\quad,\phantom{\quad}
\end{equation}
which implies that $\widehat T=P$ here. In contrast,
\begin{equation} \label{U2olaf}
\textrm{U}_\star(2): \qquad
|T\rangle \ =\ \begin{pmatrix} 
0         & 0         & \dots & 0             & |{\cal H}\rangle \\[4pt]
|0\rangle & |1\rangle & \dots & |r{-}1\rangle & \emptyset        \end{pmatrix}
\ =\ \begin{pmatrix} S_r \\[4pt] P_r \end{pmatrix} |{\cal H}\rangle
\ =:\ \widehat T\,|{\cal H}\rangle
\end{equation}
with $\emptyset\equiv 0\ 0\ \dots\ $ 
and the standard rank-$r$ projector and shift operator,
\begin{equation}
P_r \ =\ \sum_{n=0}^{r-1} |n\rangle\langle n| \undolaf
S_r \ =\ \sum_{n=r}^\infty |n{-}r\rangle\langle n| \quad,
\end{equation}
respectively. The $\textrm{U}_\star(2)$ operator~$\widehat T$ 
in~(\ref{U2olaf}) can be written as a (slightly singular) limit
of a regular expression:
\begin{equation} \label{mulimitolaf}
U(\mu)\,\widehat T \ =\ 
\begin{pmatrix} a^r \\ \mu \end{pmatrix} \,
\frac{1}{\sqrt{a^{\dagger r}a^r{+}\mu\bar\mu}}
\qquad\buildrel{\mu\to0}\over\longrightarrow\qquad
U(0)\,\widehat T \ =\ \widehat T \ =\ 
\begin{pmatrix} S_r \\ P_r \end{pmatrix}
\end{equation}
with a particular unitary transformation~$U(\mu)$.
This transformation relates the projectors smoothly as
\begin{equation}
U(\mu) \begin{pmatrix} 
\unityolaf_{\cal H} & {\bf0}_{\cal H} \\[8pt] {\bf0}_{\cal H} & P_r 
\end{pmatrix} U^\dagger(\mu) \ =\ \begin{pmatrix}
a^r \frac{1}{a^{\dagger r}a^r{+}\mu\bar\mu} a^{\dagger r} \ & \ 
a^r \frac{\bar\mu}{a^{\dagger r}a^r{+}\mu\bar\mu} \\[12pt]
\frac{\mu}{a^{\dagger r}a^r{+}\mu\bar\mu} a^{\dagger r} &
\frac{\mu\bar\mu}{a^{\dagger r}a^r{+}\mu\bar\mu} \end{pmatrix} \quad.
\end{equation}
Note that for the construction of~$P$ I can drop the square root
in~(\ref{mulimitolaf}) as effecting a basis change in im$P$ and use
$\ \widehat T=\bigl(\begin{smallmatrix} a^r \\ \mu \end{smallmatrix}\bigr)$.
For $\textrm{U}_\star(2)$, the BPS condition~(\ref{BPSTolaf}) generalizes to
\begin{equation}
\bigl(\begin{smallmatrix} a & 0 \\ 0 & a \end{smallmatrix}\bigr)\,|T\rangle\ =\
|T\rangle\,\Gamma_{\infty\times\infty} \qquad\Longleftrightarrow\qquad
a\,\widehat T \ =\ \widehat T\,\widehat\Gamma
\end{equation}
for some operator~$\widehat\Gamma$. Choosing $\widehat\Gamma=a$,
the BPS equation reduces to the holomorphicity condition
\begin{equation}
[\,a\,,\,\widehat T\,] \ =\ 0 \quad,
\end{equation}
which is indeed obeyed by the solution above. By inspection, the nonabelian
Ward solitons smoothly approach their commutative cousins for~$\theta\to0$.

\section{Moduli space dynamics}

\subsection{Manton's paradigm}

A qualitative understanding of soliton scattering can
be achieved for small relative velocity via the adiabatic or moduli-space
dynamics invented by Manton~\cite{mantono,mantonbook}. 
This approach approximates the exact scattering configuration 
of $m$ rank-one solitons by a time sequence of static $m$-lump 
solutions~$\widehat\phi(z,\bar z;\alpha)$. For the $\textrm{U}_\star(1)$
sigma model the latter are constructed from~(\ref{cohTolaf}) for $r=m$. 
Thereby one introduces a time dependence for the 
moduli~$\alpha\equiv\{\alpha_i\}$, which is determined by extremizing 
the action on the moduli space~${\cal M}_r\ni\alpha$.
Being a functional of finitely many moduli~$\alpha_i(t)$, 
this action describes the motion of a point particle in~${\cal M}_r$, 
equipped with a metric~$g_{ij}(\alpha)$ and a magnetic field~$A_i(\alpha)$. 
Hence, the scattering of $r$ slowly moving rank-one solitons is well described
by a geodesic trajectory in~${\cal M}_r$, possibly with magnetic forcing. 
Since the $\textrm{U}_\star(1)$ moduli are the spatial locations of 
the individual quasi-static lumps, the geodesic in~${\cal M}_r$ may be viewed 
as trajectories of the various lumps in the common Moyal plane, modulo 
permutation symmetry. Manton posits that
\begin{equation}
\widehat\phi(t,z,\bar z)\ \approx\ \widehat\phi(z,\bar z;\alpha(t))\ =:\ 
\phi_\alpha \quad,
\end{equation}
thus replacing dynamics for~$\widehat\phi(t,z,\bar z)$ with dynamics 
for~$\alpha(t)$. Quite generally, starting from an action of the type
\begin{equation}
S[\phi] = \int\!\D{t}\,\D^2{z}\ \bigl[ \sfracolaf12\dot\phi^2\ +\
C_\star(\phi,\phi')\,\dot\phi\ -\ W_\star(\phi,\phi') \bigr]
\qquad\textrm{with}\quad\phi'\equiv(\partial_z\phi,\partial_{\bar z}\phi)\quad,
\end{equation}
I am instructed to compute
\begin{equation}
\begin{aligned}
S_{\textrm{mod}}[\alpha] &\ :=\ S[\phi_\alpha] \\[6pt]
\ =\ &\int\!\D{t}\; \bigl[\,
\sfracolaf12 \{\smallint(\partial_\alpha\phi_\alpha)^2\}\, \dot\alpha^2 \ +\
\{\smallint C_\star(\phi_\alpha,\phi'_\alpha)\,\partial_\alpha\phi_\alpha\}\,
\dot\alpha \ -\
\smallint W_\star(\phi_\alpha,\phi'_\alpha)\,\bigr] \\[6pt]
\ =:\,&\int\!\D{t}\; \bigl[\,
\sfracolaf12 g_{\alpha\alpha}(\alpha)\, \dot\alpha^2\ +\ 
A_\alpha(\alpha)\, \dot\alpha\ -\ U(\alpha)\,\bigr]
\end{aligned}
\end{equation}
and read off the metric~$g$, magnetic field~$F=\D{A}$ and
potential~$U$ on the moduli space.

\subsection{Ward model metric}

The Ward equation~(\ref{wardeqolaf}) follows from the action~\cite{ioazak2}
\begin{equation} \label{Wardactionolaf}
S[\phi] \ =\ \sfracolaf12 \int\!\D{t}\,\D{x}\,\D{y}\;\trolaf_{\textrm{U}(n)}
\bigl[ \dot\phi^\dagger\dot\phi - 
\partial_x\phi^\dagger\partial_x\phi - 
\partial_y\phi^\dagger\partial_y\phi \bigr]
\ +\ \textrm{WZW term} \quad,
\end{equation}
both for commutative and noncommutative unitary fields.
Let me impose the space-space Moyal deformation, choose the abelian case 
($n{=}1$), pass to the operator formulation, insert the static solution
\begin{equation}
\widehat\Phi \ =\ \unityolaf\ -\ 2\,P\bigl(\{\alpha_\ell\}\bigr)
\qquad\textrm{with}\quad \trolaf P = r
\end{equation}
into $S[\Phi]$ and integrate over the Moyal plane, i.e.~perform the
trace over~$\cal H$.
Then, the gradient term in~(\ref{Wardactionolaf}) contributes with
$-\int\!\D{t}\,E[\phi_\alpha]=-8\pi r\int\!\D{t}\,1$ to $S_{\text{mod}}$
and can be dropped. More importantly, the WZW term yields 
\begin{equation}
A_\alpha \ =\ \partial_\alpha\Omega \qquad\Longrightarrow\qquad F\ =\ 0\quad,
\end{equation}
hence it too can be ignored and fails to produce a magnetic forcing
(see also~\cite{dunman})!
It remains to find the metric $g_{\alpha_i\alpha_j}(\{\alpha_\ell\})$
on the moduli space 
\begin{equation}
{\cal M}_r\ =\ \Colaf^r/S_r \ =\ 
\Colaf_\textrm{center-of-mass}\times{\cal M}_\textrm{rel}
\qquad\textrm{with}\qquad {\cal M}_\textrm{rel}\ \simeq\ \Colaf^{r-1} \quad, 
\end{equation}
which is the
configuration space of $r$~identical bosons on the Moyal plane. The result is
\begin{equation} \label{Smodolaf}
S_{\textrm{mod}} \ =\ 4\pi\theta\int\!\D{t}\;\trolaf_{\cal H} \dot P^2
\ =\ 8\pi\theta\int\!\D{t}\;\trolaf_{\textrm{im}P} \bigl(
\langle T|T\rangle^{-1}\, \langle\dot T|\unityolaf{-}P|\dot T\rangle \bigr)
\end{equation}
with \ 
$|T\rangle=\bigl(|\alpha_1\rangle,|\alpha_2\rangle,\dots,|\alpha_r\rangle\bigr)$
\ and
\begin{equation}
|\dot T\rangle \ \equiv\ \partial_t |T\rangle 
\ =\ a^\dagger |T\rangle \,\dot\Gamma\ -\ 
\sfracolaf12 |T\rangle \,(\Gamma^\dagger \Gamma)\,\dot{}
\qquad\textrm{where}\quad
\Gamma = \textrm{diag}(\{\alpha_\ell\}) \quad.
\end{equation}
It is not hard to see that the metric hiding in~(\ref{Smodolaf}) is K\"ahler,
with the K\"ahler potential~$K$ given by
\begin{equation}
\sfracolaf{1}{8\pi\theta}\,K\ =\ 
\sum_i|\alpha_i|^2\ +\ \trolaf\,\ln\bigl(\langle\alpha_i|\alpha_j\rangle\bigr)
\ =\ \trolaf\,\ln\bigl( \E^{\bar\alpha_i\alpha_j} \bigr) \quad,
\end{equation}
which makes the permutation symmetry manifest. This K\"ahler structure is
the natural one, induced from the embedding Grassmannian~Gr($r,\cal H$),
enjoys a cluster decomposition property and allows for easy separation of
the free center-of-mass motion. In the coinciding limits $\alpha_i\to\alpha_j$,
coordinate singularities appear which, however, may be removed by a gauge
transformation of~$K$ or, equivalently, by passing to permutation invariant 
coordinates (see also~\cite{liroun,haliroun,goheaspr}).

\subsection{Adiabatic two-soliton scattering}

\begin{figure}
\psfrag{$Omega$}{$g_{rr}$}
\psfrag{$r$}{$r$}
\centering
\includegraphics[height=6cm]{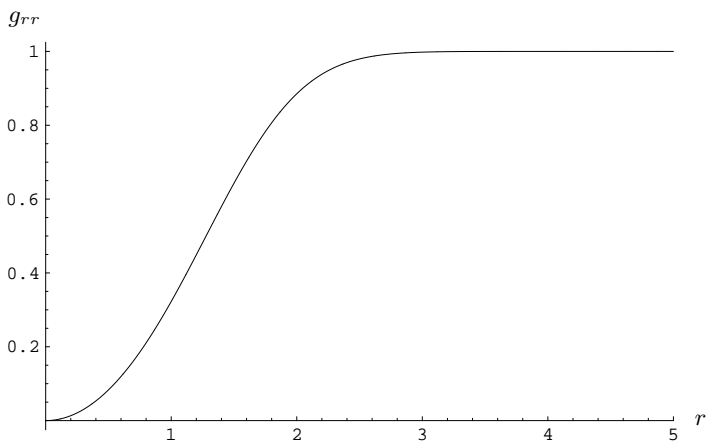}
\caption{Conformal factor of two-soliton metric}
\label{fig:olaf3}
\end{figure}

Let me be explicit for the simplest case of~$m=r=2$.
The moduli space~${\cal M}_2$ of rank-two BPS projectors is parametrized
by $\{\alpha,\beta\}\simeq\{\beta,\alpha\}\in\Colaf^2/S_2$, hence
${\cal M}_\textrm{rel}\simeq\Colaf$ but curved.
The static two-lump configuration derived from~(\ref{cohprojolaf}) reads
\begin{equation}
\Phi \ =\ \unityolaf\ -\ \frac{2}{1-|\sigma|^2}
\bigl( |\alpha\rangle\langle\alpha| + |\beta\rangle\langle\beta| 
-\sigma|\alpha\rangle\langle\beta|-\bar\sigma|\beta\rangle\langle\alpha|\bigr)
\qquad\textrm{with}\quad \sigma=\langle\alpha|\beta\rangle \quad,
\end{equation}
and the corresponding K\"ahler potential becomes~\cite{liroun}
\begin{equation}
\sfracolaf{1}{8\pi\theta}K \ =\ |\alpha|^2 + |\beta|^2 + \ln(1{-}|\sigma|^2)
\ =\ \sfracolaf12|\alpha{+}\beta|^2 + \sfracolaf12|\alpha{-}\beta|^2
+ \ln \bigl(1-\E^{-|\alpha-\beta|^2} \bigr)
\end{equation}
with center-of-mass separation. Introducing the lump distance via
$\alpha{-}\beta=r\,\E^{\I\varphi}$ and putting $\alpha{+}\beta=0$, 
the relative K\"ahler potential has the limits
\begin{equation}
\sfracolaf{1}{8\pi\theta} K_\textrm{rel} = 
\sfracolaf12 r^2 - \E^{-r^2} + O(\E^{-2r^2}) 
\undolaf
\sfracolaf{1}{8\pi\theta} K_\textrm{rel} =
\ln r^2 + \sfracolaf{1}{24}r^4 + O(r^8)
\quad,
\end{equation}
revealing asymptotic flat space for $r\to\infty$ 
but a conical singularity with an opening angle of~$4\pi$ at $r=0$.
The ensueing metric takes the conformally flat form
\begin{equation}
\D{s}^2 \ =\ 4\pi\theta\,g_{rr}(r) \bigl(\D{r}^2 + r^2 \D{\varphi}^2\bigr)
\end{equation}
with the conformal factor
\begin{equation}
g_{rr}(r) \ =\ \frac{1 - \E^{-2r^2} - 2 r^2\E^{-r^2}}{(1 - \E^{-r^2})^2}
\ =\ \frac{\sinh r^2 - r^2}{\cosh r^2 - 1} 
\ \approx\ \frac{r^2}{3} - \frac{r^6}{90} + O(r^{10}) 
\end{equation}
displayed in figure~\ref{fig:olaf3}.

\noindent
In terms of the symmetric coordinate
\begin{equation}
\rho\;\E^{\I\gamma}\ =\ \sigma\ :=\
(\alpha{-}\beta)^2 \ =\ r^2\,\E^{2\I\varphi} 
\end{equation}
the metric desingularizes,
\begin{equation}
\D{s}^2 \ =\ 4\pi\theta\,\sfracolaf{g_{rr}(\sqrt{\rho})}{4\rho}\,
\bigl(\D{\rho}^2 + \rho^2 \D{\gamma}^2\bigr) \ =\
4\pi\theta\,\bigl(\sfracolaf{1}{12}-\sfracolaf{\rho^2}{360}+O(\rho^4)\bigr)
\bigl(\D{\rho}^2 + \rho^2 \D{\gamma}^2\bigr) 
\end{equation}
which is smooth at the origin. 
Head-on scattering of two lumps corresponds to a single radial trajectory
in~${\cal M}_\textrm{rel}$, which in the smooth coordinate~$\sigma$
must pass straight through the origin.
In the `doubled coordinate'~$\alpha{-}\beta$, I then see two straight
trajectories with $90^\circ$ scattering off the singularity in the Moyal plane.
Increasing the impact factor, the scattering angle decreases from~$\frac\pi2$
to~$0$. Moreover, due to the absence of potential and magnetic field,
the scattering angle depends only on the impact parameter and not separately on
kinietc energy and angular momentum. A comparison of this moduli-space motion
with exact two-soliton dynamics has recently been performed in~\cite{klalepe}.

\section{Stability analysis}

\subsection{Fluctuation Hessian}

So far I have investigated the soliton dynamics purely on the classical level.
As a first step towards quantization, let me now turn to fluctuations around 
the classical solutions. More concretely, I shall consider perturbations of the
$2d$ static noncommutative sigma-model solitons encountered earlier. 
This task has two applications: 
First, it is relevant for the semiclassical evaluation of
the Euclidean path integral, revealing potential {\it quantum\/} instabilities
of the two-dimensional model. Second, it yields the (infinitesimal) time
evolution of fluctuations around the static multi-soliton in the time-extended
three-dimensional theory, indicating {\it classical\/} instabilities if they
are present. More concretely, any static perturbation of a classical
configuration can be taken as (part of the) Cauchy data for a classical time
evolution, and any negative eigenvalue of the quadratic fluctuation operator
will give rise to an exponential runaway behavior, at least within the linear
response regime. Furthermore, fluctuation zero modes are expected to belong
to moduli perturbations of the classical configuration under consideration.
The current knowledge on the effect of quantum fluctuations is summarized in
\cite{zakbook}.
For a linear stability analysis, I must study the $\textrm{U}_\star(n)$ energy
functional~(\ref{Eolaf}) for a perturbation~$\phi$ of a background~$\Phi$,
\begin{equation}
E[\Phi{+}\phi] \ =\ E[\Phi] \ +\ \delta E[\Phi,\phi] \ +\ \delta^2 E[\Phi,\phi]
\ +\ \dots \quad,
\end{equation}
where the $\phi$-linear term $\delta E$ vanishes for classical backgrounds, and
\begin{equation}
\delta^2 E[\Phi,\phi] \ =\ 2\pi\,\trolaf\bigl\{
\phi^\dagger\,\Delta\phi - \phi^\dagger\,(\Phi\Delta\Phi^\dagger)\;\phi \bigr\}
\ =:\ 2\pi\,\trolaf\bigl\{ \phi^\dagger\,H\,\phi \bigr\}
\end{equation}    
defines the Hessian operator~$H[\Phi]$ which acts in the space of fluctuations
$\phi$. For a given static soliton~$\Phi$, the goal is to determine the 
spectrum of~$H$, at least the negative part and the zero modes.
To this end, a decomposition of~$\{\phi\}$ into $H$-invariant subspaces is
essential. A natural segmentation is
\begin{equation}
\phi\ =\ \biggl( \begin{matrix} \vphantom{\bigm|}
\phi_{\textrm{im}P} & \vline & \phi_{\textrm{Gr}P} \\
\hline 
\phi_{\textrm{Gr}P} & \vline & \phi_{\textrm{ker}P}
\end{matrix} \biggr) \qquad\textrm{on}\quad
{\cal H}=\textrm{im}P\oplus\textrm{ker}P \quad.
\end{equation}
Here $\phi_{\textrm{Gr}P}$ is hermitian and keeps me inside the Grassmannian
of~$\Phi$, while $\phi_{\textrm{im}P}$ and $\phi_{\textrm{ker}P}$ are
anti-hermitian and lead away from the Grassmannian.
Even though this structure is not $H$-invariant, it decomposes the energy,
\begin{equation}
\delta^2 E[\Phi,\phi] \ =\ 
\delta^2 E[\Phi,\phi_{\textrm{im}P}{+}\phi_{\textrm{ker}P}]\ +\
\delta^2 E[\Phi,\phi_{\textrm{Gr}P}] \quad.
\end{equation}
Without further assumptions about~$\Phi$ it is difficult to identify
$H$-invariant subspaces. Let me adopt the basis~(\ref{Nbaseolaf}) in~$\cal H$.
Then, for backgrounds {\it diagonal\/} in this basis, an $H$-invariant 
decomposition is
\begin{equation} \label{decompolaf}
\phi \ =\ \sum_{k=0}^\infty \phi^{(k)} \quad,
\end{equation}
where $\phi^{(k)}$ denotes the $k$th diagonal plus its transpose.

\subsection{Diagonal $\textrm{U}_\star(1)$ soliton: fluctuation spectrum}

\begin{figure}
\psfrag{phi=}{$\phi\ =\ $}
\psfrag{im P}{$\scriptstyle{\text{im}P}$}
\psfrag{ker P}{$\scriptstyle{\text{ker}P}$}
\psfrag{d Gr(P)}{$\scriptstyle{\text{d\,Gr($P$)}}$}
\psfrag{u(im P)}{$\scriptstyle{u(\text{im}P)}$}
\psfrag{u(ker P)}{$\scriptstyle{u(\text{ker}P)}$}
\centering
\includegraphics[height=5cm]{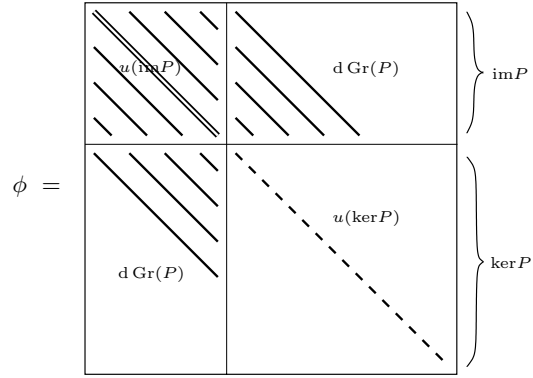}
\caption{Decomposition of perturbation around $\Phi_4$}
\label{fig:olaf4}
\end{figure}

\begin{figure}
\psfrag{$k$}{$\scriptstyle{k}$}
\psfrag{$lambda$}{$\scriptstyle{\lambda}$}
\centering
\includegraphics[height=5cm]{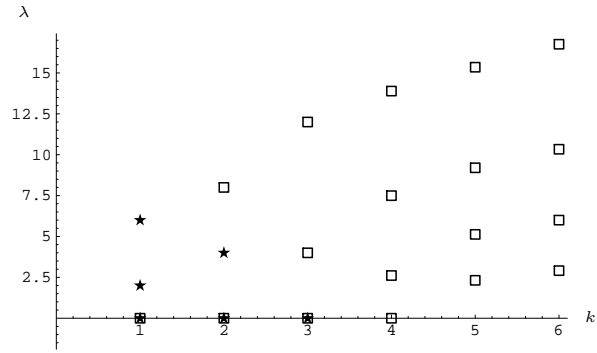}
\caption{Discrete spectrum of $H$ for $\Phi_4$}
\label{fig:olaf5}
\end{figure}

\begin{figure}
\psfrag{$k$}{$\scriptstyle{k}$}
\psfrag{$lambda$}{$\scriptstyle{\lambda}$}
\centering
\includegraphics[height=5cm]{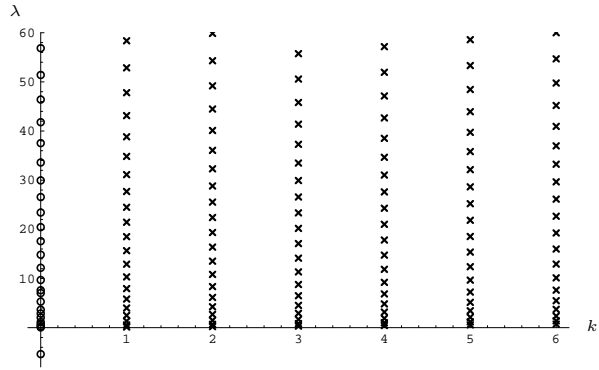}
\caption{Continuous spectrum of $H$ for $\Phi_4$}
\label{fig:olaf6}
\end{figure}

Once more I specialize to the abelian sigma model, where each static soliton
is essentially a coherent-state projector~(\ref{cohprojolaf}) labelled by
$r$ complex numbers~$\alpha_i$. Although all these backgrounds (for fixed~$r$)
are degenerate in energy, their fluctuation spectra differ unless related
by ISO(2) rigid motion in the Moyal plane. Presently, the fluctuation analysis
is technically feasible only for the special backgrounds where all $\alpha_i$
coalesce. Translating the common value to the origin, this amounts to the
diagonal abelian background
\begin{equation}
\Phi_r \ =\ \unityolaf\ -\ 2\sum_{n=0}^{r-1} |n\rangle\langle n| \ =\ 
\textrm{diag}(\underbrace{-1,-1,\dots,-1}_{r\ \textrm{times}},+1,+1,\dots)\quad.
\end{equation}
In this case, the decomposition~(\ref{decompolaf}) applies and yields three
qualitatively different types of fluctuation subspaces carrying the following
characteristic spectra of~$H$:
\begin{equation}
\begin{aligned}
k>r \qquad & \left\{ \ \begin{matrix}
\textrm{spec}(H^{(k)}_{\textrm{Gr}P})=\{0<\lambda_1<\dots<\lambda_r\} \\[8pt]
\textrm{spec}(H^{(k)}_{\textrm{im}P})=\emptyset \hfill \\[8pt]
\textrm{spec}(H^{(k)}_{\textrm{ker}P})=\Rolaf_+ \hfill \end{matrix}
\ \right\} \quad \textrm{`very off-diagonal'} \\[12pt]
1\le k\le r \ & \left\{ \ \begin{matrix}
\textrm{spec}(H^{(k)}_{\textrm{Gr}P})
=\{0=\lambda_1<\lambda_2<\dots<\lambda_k\} \hfill \\[8pt]
\textrm{spec}(H^{(k)}_{\textrm{im}P})
=\{0=\lambda_{k+1}<\lambda_{k+2}<\dots<\lambda_r\} \\[8pt]
\textrm{spec}(H^{(k)}_{\textrm{ker}P})
=\Rolaf_+ \hfill \end{matrix}
\ \right\} \quad
\textrm{\lower8pt\vbox{\flushleft `slightly \break off-diagonal'}} \\[12pt]
k=0 \qquad & \left\{ \ \begin{matrix}
\textrm{spec}(H^{(0)}_{\textrm{Gr}P})=\emptyset \hfill \\[8pt]
\textrm{spec}(H^{(0)}_{\textrm{im}P+\textrm{ker}P})
=\Rolaf_{\ge0} \cup\{\lambda_- <0\} 
\end{matrix} \ \right\} \quad \textrm{`diagonal'}
\end{aligned}
\end{equation}
These findings are visualized for $r{=}4$ in figure~\ref{fig:olaf4},
with the following legend: \\[8pt]
\begin{tabular}{lclcl} 
double line & $\quad\widehat=\quad$ & negative eigenvalue & 
$\qquad\longrightarrow\qquad$ & single instability \\
solid segment & $\quad\widehat=\quad$ & zero eigenvalue & 
$\qquad\longrightarrow\qquad$ & $(2r{-}1)_\Colaf$ moduli \\
dashed line & $\quad\widehat=\quad$ & admissible zero mode & 
$\qquad\longrightarrow\qquad$ & phase modulus 
\end{tabular} \\[8pt]
Figures~\ref{fig:olaf5} and \ref{fig:olaf6} show a numerical
spectrum of the~$H^{(k)}$ with cut-off size 30, 
also for the background~$\Phi_4$. Here, the legend is: \\[8pt]
\phantom{XXX}
\begin{tabular}{lclcl} 
boxes   & $\quad\widehat=\quad$ & Gr($P$) eigenvalues & 
$\qquad\longrightarrow\qquad$ & $\#=\textrm{min}(r,k)$ \\
stars   & $\quad\widehat=\quad$ & im$P$ eigenvalues ($k{\neq}0$) & 
$\qquad\longrightarrow\qquad$ & $\#=\textrm{max}(r{-}k,0)$ \\
crosses & $\quad\widehat=\quad$ & ker$P$ modes ($k{\neq}0$) & 
$\qquad\longrightarrow\qquad$ & $\Rolaf_+$ continuum \\
circles & $\quad\widehat=\quad$ & diagonal modes ($k{=}0$) &
$\qquad\longrightarrow\qquad$ & $\Rolaf_{\ge0}\cup\{\lambda_-\}$ 
\end{tabular} 

\subsection{Single negative eigenvalue}

The numerical analysis for abelian diagonal backgrounds~$\Phi_r$ revealed 
a single negative eigenvalue~$\lambda_-$ among the diagonal fluctuations.
It is found by diagonalizing the $k{=}0$ part of the Hessian,
\begin{equation}
(H^{(0)}_{m\ell}) \ =\ \begin{pmatrix}
\phantom{-}1 &           -1 \\[2pt]
          -1 & \phantom{-}3 &           -2 \\[2pt]
             &           -2 & \phantom{-}5 & -3 \\[-2pt]
             &              &           -3 & \ddots & \ddots \\[-2pt]
             & & & \ddots & 2r{-}3 & -r{+}1 & & & \\[2pt]
             & & & & -r{+}1 & {\bf-1} & -r & & \\[2pt]
             & & & & & -r & {\bf+1} & -r{-}1 & \\[-2pt]
             & & & & & & -r{-}1 & 2r{+}3 & \ddots \\[-2pt]
             & & & & & & & \ddots & \ddots
             \end{pmatrix} \quad,
\end{equation}
where I have emphasized in boldface the entries modified by the background.
The result is indeed that
\begin{equation}
\textrm{spec}(H^{(0)})\ =\ \{\lambda_-\}\cup[0,\infty) \quad,
\end{equation}
where $\lambda_-$ is computed as the unique negative zero of the determinant
\begin{equation}
\left|\begin{matrix}
I_{r-1,r-1}(\lambda)-\frac{1}{2r}  & I_{r-1,r}(\lambda) \\[8pt]
I_{r,r-1}(\lambda) & I_{r,r}(\lambda)-\frac{1}{2r} \end{matrix}\right| 
\qquad\textrm{with}\quad
I_{k,l}(\lambda) := 
\int_0^{\infty} \frac{\E^{-x}\,\D x}{x-\lambda}\; L_k(x)\,L_l(x)
\end{equation}
being variants of the integral logarithm.
The $r$ complex zero eigenvalues of $H_{\textrm{Gr}P}$ arise from
turning on the location moduli~$\alpha_i$ of~(\ref{cohTolaf}),
while the $r{-}1$ complex zero eigenvalues of $H_{\textrm{im}P}$
point at non-Grassmannian classical solutions. Since $H^{(0)}$ is not
non-negative, $\delta^2 E[\Phi,\phi^{(0)}]$ may vanish even if
$H^{(0)}\phi^{(0)}\neq0$.

\subsection{Instability in unitary sigma model}

The fluctuations~$\phi_{\textrm{Gr}P}$ are tangent to 
Gr$P\equiv$ Gr$(r,\cal H)$ and cannot lower the energy, 
as the BPS argument~(\ref{BPSEolaf}) had assured me from the beginning. 
Therefore, all solitons of Grassmannian sigma models are stable.
On the other hand, an unstable mode of~$H$ occurred in im$P\oplus$ker$P$,
indicating a possibility to continuously lower the energy $E=8\pi r$ of
$\Phi_r$ along a path starting perpendicular to~Gr$P$.
Indeed, there exists a general argument for {\it any\/} static soliton
$\Phi=\unityolaf{-}2P$ inside the unitary sigma model, commutative or 
noncommutative. It goes as follows. Given a projector inclusion
$\widetilde P\subset P$ (including $\widetilde P=0$), i.e.~a `smaller'
projector~$\widetilde P$ of rank~$\widetilde r<r$. 
Then, the path~\cite{zakbook}
\begin{equation}
\Phi(s) \ =\ \E^{\I\,s\,(P-\widetilde P)} (\unityolaf{-}2P) 
\ =\ \unityolaf -(1{+}\E^{\I\,s})P -(1{-}\E^{\I\,s})\widetilde P
\end{equation}
\begin{equation}
\textrm{connecting}\qquad \Phi(0)=\Phi=\unityolaf{-}2P \qquad
\textrm{to}\qquad \Phi(\pi)=\widetilde\Phi=\unityolaf{-}2\widetilde P
\end{equation}
interpolates between static solitons in different Grassmannians 
inside~U($\cal H$). Please note that the tangent vector
$(\partial_s\Phi)(0)=-\I(P{-}\widetilde P)$ is {\it not\/} an eigenmode
of the Hessian. A quick calculation gives the energy along the path,
\begin{equation}
\sfracolaf{1}{8\pi} E[\Phi(s)] 
\ =\ \sfracolaf{r+\widetilde r}{2}\ +\ \sfracolaf{r-\widetilde r}{2}\cos s
\ =\ r\,\cos^2\sfracolaf{s}{2}\ +\ \widetilde r\,\sin^2\sfracolaf{s}{2}\quad.
\end{equation}
For nonabelian noncommutative solitons the argument persists, with the
topological charges $Q$ and $\widetilde Q$ replacing $r$ and $\widetilde r$.
Therefore, all solitons in unitary sigma models eventually decay to the
`vacua' $Q=0$, which belong to the constant (nonabelian) projectors.

\section{$d=1{+}1$ sine-Gordon solitons}

\subsection{Reduction to $d=(1{+}1)_\theta$: instantons}

\begin{figure}
\centering
\includegraphics[height=4cm]{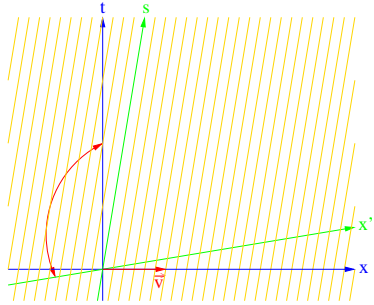}
\caption{Action of reduction $\partial_s=0$}
\label{fig:olaf7}
\end{figure}

\begin{figure}
\centering
\includegraphics[height=5cm]{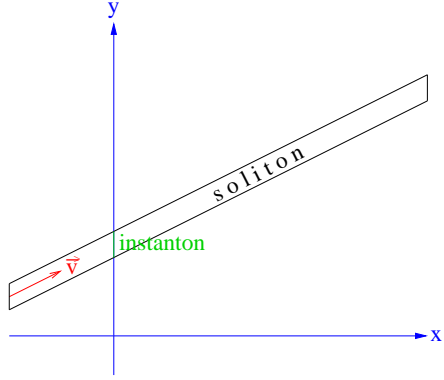}
\caption{$x{=}0$ instanton snapshot of soliton}
\label{fig:olaf8}
\end{figure}

In the remaining part of this lecture I look at the reduction from 
$1{+}2$ to $1{+}1$ dimensions, with the goal to generate new
noncommutative solitons. However, naive reduction of the Ward solitons
is not possible. Due to shape invariance, $\partial_s=0$, the one-soliton 
sector is already two-dimensional (in the rest-frame) but with Euclidean 
signature:
\begin{equation} \label{redolaf}
\partial_s=0 \qquad\leftrightarrow\qquad
\partial_u + \mu\bar\mu\,\partial_v -(\mu{+}\bar\mu)\,\partial_x = 0
\qquad\leftrightarrow\qquad
\partial_x = \nu\,\partial_w + \bar\nu\,\partial_{\bar w} \quad,
\end{equation}
hence I cannot simply put $\partial_x=0$ without killing the soliton entirely.
Instead, the $x$~dependence may be eliminated by taking the 
snapshot~$\phi(x{=}0,y,t)$. Then, $\partial_s=0$ maps the remaining $ty$~plane
to the $w\bar w$~plane as illustrated in figure~\ref{fig:olaf7}.
Because for $\velolaf_x{\neq}0$ the soliton worldline pierces the $xy$~plane
as shown in figure~\ref{fig:olaf8},
the $x{=}0$~slice of the soliton is just an {\it instanton\/}!

\subsection{$d=1{+}1$ sigma model metric}

Due to the $x$-derivatives in the Ward equation~(\ref{wardeqolaf})
the snapshot $\phi(x{=}0,y,t)$ will not satisfy this equation.
Using~(\ref{redolaf}) I find that instead it obeys the equation
\begin{equation}
(1{-}\sfracolaf{\mu}{\bar\mu})\,\partial_w(\phi^\dagger\partial_{\bar w}\phi) -
(1{-}\sfracolaf{\bar\mu}{\mu})\,\partial_{\bar w}(\phi^\dagger\partial_w \phi)
\ =\ 0 \quad,
\end{equation}
which is an extended sigma-model equation in $1{+}1$ dimensions due to
$(w,\bar w)\sim(t,y)$. Comparison with
\begin{equation}
(h^{(ij)}+b^{[ij]})\,\partial_i(\phi^\dagger\partial_j \phi) \ =\ 0
\qquad\textrm{for}\quad i,j\in\{t,y\}
\end{equation}
yields the metric
\begin{equation}
\begin{pmatrix}
\ h^{tt}\ & \ h^{ty}\ \\[12pt]
\ h^{yt}\ & \ h^{yy}\ \end{pmatrix} \ =\
\frac{\mu\bar\mu}{(\mu{+}\bar\mu)^2}
\begin{pmatrix}
|\frac{1}{\mu}{-}\mu|^2 & \ |\frac{1}{\mu}|^2 - |\mu|^2 \\[12pt]
|\frac{1}{\mu}|^2 - |\mu|^2 & \ |\frac{1}{\mu}{+}\mu|^2
\end{pmatrix}
\end{equation}
and the magnetic field
\begin{equation}
\begin{pmatrix}
\ b^{tt}\ & \ b^{ty}\ \\[6pt]
\ b^{yt}\ & \ b^{yy}\ \end{pmatrix} \ =\
\begin{pmatrix}
\phantom{-}0 \ & \ 1 \ \\[6pt]
          -1 \ & \ 0 \ \end{pmatrix} \quad.
\end{equation}
The notation suggests a Minkowski signature, but a short computation says that
\begin{equation}
\det(h^{ij})\ =\ 
\bigl(\sfracolaf{\textrm{Im}\,\mu}{\textrm{Re}\,\mu}\bigr)^2\ \ge\ 0 \quad,
\end{equation}
hence the metric is Euclidean! Indeed, this very fact permits the Fock-space
realization of the Moyal deformation, which follows.

\subsection{Moyal deformation in $d=1{+}1$}

In the present case I have no choice but to employ the time-space deformation
\begin{equation}
[t,y]_\star = \I\theta \qquad\Longrightarrow\qquad 
[w,\bar{w}]_\star = 2\theta \qquad\textrm{for}\quad
\mu\notin\Rolaf \ \textrm{or}\ \I\,\Rolaf \quad.
\end{equation}
As before, I realize this algebra via the Moyal-Weyl correspondence
\begin{equation}
w\ \leftrightarrow\ \sqrt{2\theta}\,c \qquad\textrm{such that}\qquad
[\,c\,,\,c^\dagger]\ =\ \unityolaf 
\end{equation}
on the standard Fock space~$\cal H$.
In this way, the moving $\textrm{U}_\star(1)$ soliton~(\ref{onesololaf})
becomes a gaussian instanton in the $d=1{+}1\ \textrm{U}_\star(1)$ sigma model,
after reexpressing $w=w(y,t)$.  The only exception occurs for 
$\velolaf_x=0 \ (\Leftrightarrow \mu\in\I\,\Rolaf)$, 
i.e.~motion in $y$~direction only, because (\ref{nctypesolaf}) then implies
that $[w,\bar{w}]_\star=0$. In fact, (\ref{restolaf})~shows (for $x{=}0$) 
that $\bar w\sim w$ in this case, the rest frame degenerates to one dimension
and there is no room left for a Heisenberg algebra.\quad{\lower8pt\hbox{
$\buildrel{\bullet\,\bullet}\over{\buildrel{\angle\ }\over\frown}$} }

\subsection{Reduction to $d=(1{+}1)_\theta$: solitons}

So far, my attempts to construct noncommutative solitons in $1{+}1$ dimensions
by reducing such solitons in a $d{=}1{+}2$ model have failed.
The lesson to learn is that the dimensional reduction must occur along 
a spatial symmetry direction of the $d{=}1{+}2$ configuration, i.e.~along
its worldvolume. In other words, the starting configuration should be
spatially extended, or a $d{=}1{+}2$ noncommutative wave! Luckily, 
such wave solutions exist in the nonabelian Ward model~\cite{leeseo,bielingo}.
Let me warm up with the commutative case and the sigma-model group of~U(2).
The Ward-model wave solutions~$\Phi(u,v,x)$ dimensionally reduce 
to $d{=}1{+}1$ WZW solitons~$g(u,v)$ via
\begin{equation} \label{phiredolaf}
\Phi(u,v,x) \ =\ {\cal E}\, \E^{\I\alpha\,x\,\sigma_1}\;g(u,v)\,
\E^{-\I\alpha\,x\,\sigma_1}\,{\cal E}^\dagger
\qquad\textrm{for}\quad g(u,v)\in \textrm{U(2)}
\end{equation}
and a constant $2{\times}2$ matrix~$\cal E$.
The Ward equation for~$\Phi$ descends to 
\begin{equation} \label{wardredolaf}
\partial_v ( g^\dagger \partial_u g )\ +\ 
\alpha^2 (\sigma_1 g^\dagger \sigma_1 g - g^\dagger \sigma_1 g\,\sigma_1)
\ =\ 0 \quad.
\end{equation}
In a second step, I algebraically reduce $g$ from U(2) to being U(1)-valued,
allowing for an angle parametrization,
\begin{equation}
g \ =\ \E^{\frac{\I}{2}\sigma_3 \phi} \quad.
\end{equation}
The algebra of the Pauli matrices then simplifies (\ref{wardredolaf}) to
\begin{equation}
\partial_v \partial_u \phi + 4\alpha^2 \sin\phi \ =\ 0
\end{equation}
which is nothing but the familiar sine-Gordon equation!

\subsection{Integrable noncommutative sine-Gordon model}

Now I introduce the time-space Moyal deformation
\begin{equation}
[t,y]_\star\ =\ \I\theta
\qquad\Longleftrightarrow\qquad 
[u,v]_\star\ =\ -\sfracolaf{\I}{2}\,\theta \quad.
\end{equation}
The sine-Gordon kink must move in the $y$ direction, which (we have learned)
forbids a Heisenberg algebra (note the $\I$ above). Thus, no Fock-space
formulation exists and I must content myself with the star product.
Recalling the dimensional reduction (\ref{phiredolaf}) and~(\ref{wardredolaf})
I must now solve
\begin{equation} \label{ncwardredolaf}
\partial_v(g^\dagger\star\partial_u g)\ +\ 
\alpha^2(\sigma_1 g^\dagger\star\sigma_1 g-g^\dagger\sigma_1\star g\,\sigma_1)
\ =\ 0 \quad.
\end{equation}
The algebraic reduction U(2) $\to$ U(1) turns out to be too restrictive.
In the commutative case, the overall U(1) phase factor
$\E^{\frac{\I}{2}\unityolaf\,\rho}$ of~$g$ decouples in~(\ref{wardredolaf}),
so I could have started directly with $g\in$ SU(2) instead.
In the noncommutative case, in contrast, this does not happen, and I am forced
to begin with~$\textrm{U}_\star(2)$. Thus, I should not prematurely drop the 
overall phase and algebraically reduce~$g$ to
$\textrm{U}_\star(1)\times\textrm{U}_\star(1)$,
\begin{equation}
g(u,v)\ =\ 
\E_\star^{\frac{\I}{2}\unityolaf\,\rho(u,v)}\star
\E_\star^{\frac{\I}{2}\sigma_3\,\varphi(u,v)} \quad.
\end{equation}
With this, the $2{\times}2$ matrix equation~(\ref{ncwardredolaf}) turns into
the scalar pair
\begin{equation} \label{ncsgolaf}
\begin{aligned}
\partial_v\bigl(\E_\star^{-\frac{\I}2\varphi}\star
\partial_u\E_\star^{\frac{\I}2\varphi}\bigr)\ +\
2\I\alpha^2\sin_\star\!\varphi \ =\ 
-\partial_v\bigl[\E_\star^{-\frac{\I}2\varphi}\star 
R\star\E_\star^{\frac{\I}2\varphi} \bigr] \\[6pt]
\partial_v\bigl(\E_\star^{\frac{\I}2\varphi}\star
\partial_u\E_\star^{-\frac{\I}2\varphi} \bigr)\ -\
2\I\alpha^2\sin_\star\!\varphi \ =\ 
-\partial_v\bigl[\E_\star^{\frac{\I}2\varphi}\star 
R\star\E_\star^{-\frac{\I}2\varphi} 
\bigr]
\end{aligned}
\end{equation}
with the abbreviation
\begin{equation}
R\ =\ \E_\star^{-\frac{\I}{2}\rho}\star\partial_u\E_\star^{\frac{\I}{2}\rho}
\end{equation}
carrying the second angle~$\rho$. For me, (\ref{ncsgolaf}) are the
noncommutative sine-Gordon (NCSG) equations.
As a check, take the limit $\theta\to0$, which indeed yields
\begin{equation}
\partial_v\,\partial_u\,\rho \ =\ 0 \undolaf
\partial_v\,\partial_u\,\varphi\ +\ 4\alpha^2\sin\varphi \ =\ 0 \quad.
\end{equation}

\subsection{Noncommutative sine-Gordon kinks}

As an application I'd like to construct the deformed multi-kink solutions
to the NCSG equations~(\ref{ncsgolaf}), e.g.~via the associated linear system.
First, consider the one-kink configuration, which obtains from the wave
solution of the $\textrm{U}_\star(2)$ Ward model by choosing
\begin{equation}
x=0 \qquad\textrm{as well as}\qquad
\mu\ =\ \I\,p\,\in\I\,\Rolaf \qquad\Longrightarrow\qquad \nu=1 \quad.
\end{equation}
Consequently, the co-moving coordinate becomes
\begin{equation}
w\ =\ \bar{\mu}u+\sfracolaf{1}{\bar{\mu}}v\ =\ 
-\I\,(p\,u+\sfracolaf{1}{p}v)\ =\ 
-\I\sfracolaf{y-\velolaf t}{\sqrt{1-\velolaf^2}}\ =:\ -\I\,\eta \quad.
\end{equation}
The BPS solution of the reduced Ward equation~(\ref{ncwardredolaf}) is
\begin{equation}
g\ =\ \sigma_3(\unityolaf{-}2P) \qquad\textrm{with projector}\qquad
P\ =\ T\star\sfracolaf{1}{T^\dagger\star T}\star T^\dagger \quad,
\end{equation}
where the $2{\times}1$ matrix function $T(\eta)$ is subject to
\begin{equation}
(\partial_\eta + \alpha\,\sigma_3)\,T(\eta) \ =\ 0 \quad.
\end{equation}
Modulo adjusting the integration constant and (irrelevant) scaling factor,
the general solution reads
\begin{equation}
\begin{aligned}
T &\ =\ \biggl(\begin{matrix}
\E^{-\alpha\eta}\  \\[6pt] \I\,\E^{\alpha\eta} \end{matrix}\biggr)
\qquad\Longrightarrow\qquad
P \ =\ \frac{1}{2\,\cosh 2\alpha\eta} \biggl(\begin{matrix}
\E^{-2\alpha\eta} & -\I \\[6pt] \I & \E^{+2\alpha\eta} \end{matrix}\biggr)
\quad, \\[12pt] 
g &\ =\ \Biggl(\begin{matrix} 
\tanh 2\alpha\eta & \ \frac{\I}{\cosh 2\alpha\eta}\\[8pt]
\frac{\I}{\cosh 2\alpha\eta} & \ \tanh 2\alpha\eta \end{matrix}\Biggr)
\ \buildrel{!}\over{\ =\ }\
{\cal E}\,\Biggl(\begin{matrix}
\E_\star^{\frac{\I}{2}\,\rho}\star\E_\star^{\frac{\I}{2}\,\varphi} & 0 \\[4pt]
0 & \E_\star^{\frac{\I}{2}\,\rho}\star\E_\star^{-\frac{\I}{2}\,\varphi}
\end{matrix}\Biggr)\,{\cal E}^\dagger \quad.
\end{aligned}
\end{equation}

\subsection{One-kink configuration}

Since the expressions above depend on $u$ and~$v$ only in the rest-frame
combination~$\eta$, it is clear that the deformation becomes irrelevant here,
and the one-kink sector is commutative, effectively $\theta=0$ and $\rho=0$.
With ${\cal E}=\sfracolaf{1}{\sqrt{2}} \bigl( \begin{smallmatrix}
1 & -1 \\ 1 & \phantom{-}1 \end{smallmatrix} \bigr)$ the latest equation
is solved by
\begin{equation}
\cos\sfracolaf{\varphi}{2} = \tanh 2\alpha\eta \undolaf
\sin\sfracolaf{\varphi}{2} = \sfracolaf{1}{\cosh 2\alpha\eta}
\qquad\Longrightarrow\qquad
\tan\sfracolaf{\varphi}{4} = \E^{-2\alpha\eta}
\end{equation}
which is precisely the standard sine-Gordon kink with velocity
$\velolaf=\sfracolaf{1-p^2}{1+p^2}$.
With hindsight this was to be expected, since a one-soliton configuration
in $1{+}1$ dimensions depends on a single (real) co-moving coordinate.
The deformation should reappear, however, in multi-soliton solutions.
For instance, breather and two-soliton configurations seem to get deformed 
since pairs of rest-frame coordinates are subject to
\begin{equation}
[\eta_i\,,\eta_k]_\star\ =\ -\I\,\theta\,(\velolaf_i{-}\velolaf_k)\big/
\sqrt{\vphantom{|}\smash{(1{-}\velolaf_i^2)(1{-}\velolaf_k^2)}}
\quad.
\end{equation}

\subsection{Tree-level scattering of elementary quanta}

Finally, it is of interest to investigate the quantum structure of
noncommutative integrable theories, i.e.~take into account the
field excitations above the classical configurations. In my noncommutative
sine-Gordon model~(\ref{ncsgolaf}) the elementary quanta are $\varphi$
and~$\rho$, and the Feynman rules for their scattering do get Moyal deformed.
For illustrative purposes I concentrate on the 
$\varphi\varphi\to \varphi\varphi$ scattering amplitude in the vacuum sector.
The kinematics of this process is
\begin{equation}
k_1=(E,p)\ ,\quad
k_2=(E,-p)\ ,\quad
k_3=(-E,p)\ ,\quad
k_4=(-E,-p)\quad,
\end{equation}
subject to the mass-shell condition $\ E^2-p^2=4\alpha^2$.
The action (which I did not present here) is non-polynomial; 
it contains
\begin{equation}
\langle\varphi\varphi\rho\rangle \quad,\qquad
\langle\rho\rho\rho\rangle\quad,\qquad
\langle\varphi\varphi\varphi\varphi\rangle \quad,\qquad
\langle\varphi\varphi\rho\rho\rangle \quad,\qquad
\langle\rho\rho\rho\rho\rangle
\end{equation}
as elementary three- and four-point interaction vertices.
Denoting $\varphi$ propagators by solid lines and $\rho$
propagators by dashed ones, there are the following four
contributions to the $\varphi\varphi\to \varphi\varphi$ amplitude
at tree level:
\\[-12pt]
\begin{center}\begin{tabular}{rclrcl}
\parbox{1.9cm}{\includegraphics[width=1.9cm]{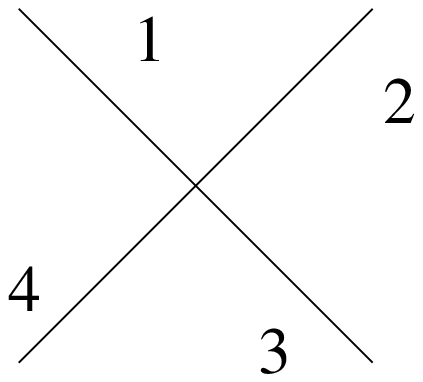}}
&=&$2\I\alpha^2 \cos^2 (\theta Ep)$ \qquad\qquad
\parbox{1.9cm}{\includegraphics[width=1.9cm]{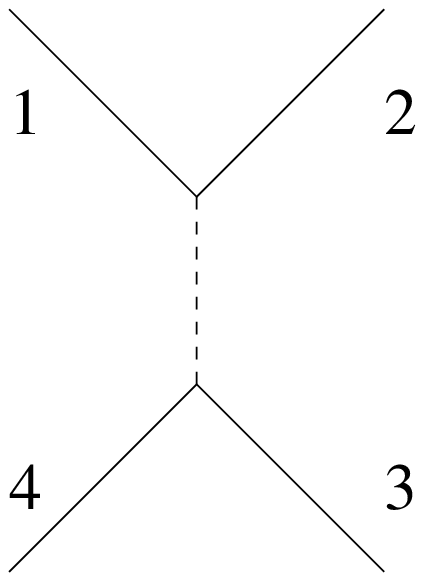}}
$=-\frac{\I}{2}p^2\sin^2 (\theta Ep)$ \\[36pt]
\parbox{2.4cm}{\includegraphics[width=2.4cm]{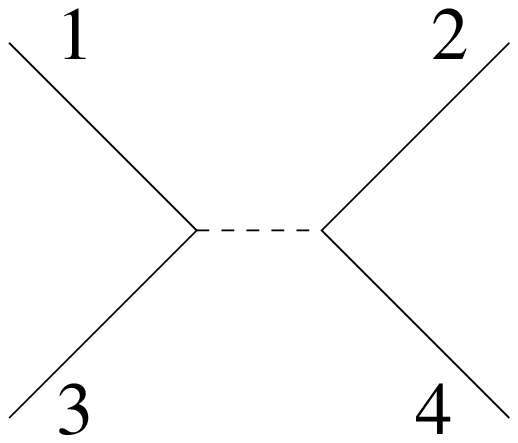}}
&=&$ \frac{\I}{2}E^2\sin^2 (\theta Ep)$ \qquad\qquad
\parbox{2.4cm}{\includegraphics[width=2.4cm]{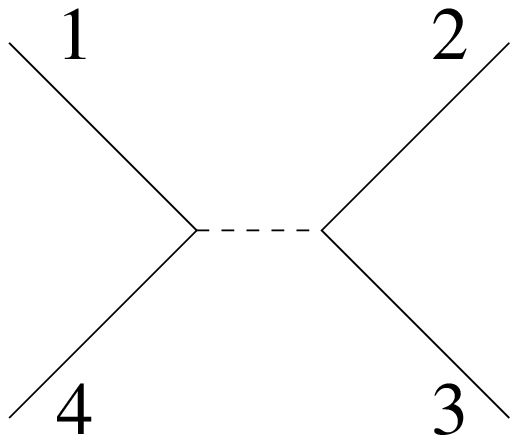}}
$=0$ \quad.
\end{tabular}\\ \end{center}
Taken together this means that
\begin{equation}
A_{\varphi\varphi\to\varphi\varphi}\ =\ 2\I\alpha^2
\end{equation}
is causal. I can show that all other $2\to2$ tree amplitudes vanish.
Hence, any $\theta$~dependence seems to cancel in the tree-level S-matrix!
Furthermore, it can be established that there is no tree-level particle
production in this model, just like in the commutative case.
Although at tree-level I still probe only the classical structure of
the theory, the absence of a deformation until this point is conspicuous:
Could it be that the time-space noncommutativity in the sine-Gordon system
is a fake, to be undone by a field redefinition?
With this provoking question I close the lecture.

%
%


\end{document}